\RequirePackage{fix-cm}
\documentclass[smallcondensed,natbib]{svjour3}     
\smartqed  
\usepackage{graphicx}
\usepackage{amssymb}
\usepackage{amsmath}
\usepackage{epstopdf}
\usepackage{color}
\usepackage{ulem}
\newcommand{\G}{\Gamma} 
\newcommand{\g}{\gamma}
\newcommand{\R}{\mathcal {R}}

\newcommand{\Ah}{A_{\rm half}}  
\newcommand{\Pm}{P_{\rm max}}  
\newcommand{\al}{\alpha}
\newcommand{\de}{\delta}
\newcommand{\m}{\tilde{m}}
\newcommand{\chis}{\tilde{\chi}}
\newcommand{\chiz}{\tilde{\psi}}  
\newcommand{\als}{\tilde{\alpha}}
\newcommand{\des}{\tilde{\delta}}
\newcommand{\rs}{r}
\newcommand{\ka}{\kappa}
\newcommand{\kas}{\tilde{\kappa}}
\newcommand{\s}{\sigma}
\newcommand{\uu}{u}
\newcommand{\lam}{\lambda}

\newcommand{\Lam}{\Lambda}
\newcommand{\emax}{\epsilon_{\rm max}}
\newcommand{\emin}{\epsilon_{\rm min}}

\begin{document}

\title{A hybrid model for the population dynamics of periodical cicadas}


\author{Jonathan Machta          \and
        Julie C Blackwood   \and
        Andrew Noble       \and
        Andrew M Liebhold     \and
        Alan Hastings
}


\institute{J. Machta \at
              Department of Physics, University of Massachusetts, Amherst, MA 01003, USA and\\
              Santa Fe Institute, 1399 Hyde Park Road, Santa Fe, New Mexico 87501, USA\\
              \email{machta@physics.umass.edu}           
           \and
           J.C. Blackwood \at
                Department of Mathematics and Statistics, Williams College, Williamstown, MA 01267, USA
           \and
           A.E. Noble \at
                Department of Environmental Science and Policy, University of California, Davis, CA 95616, USA
           \and
           A.M. Liebhold \at
                Northern Research Station, U.S. Forest Service, Morgantown, WV, 26505, USA
           \and
           A. Hastings   \at
                Department of Environmental Science and Policy, University of California, Davis, CA 95616, USA
}

\date{Received: date / Accepted: date}

\begin{abstract}

In addition to their unusually long life cycle, periodical cicadas, {\it Magicicada} spp., provide an exceptional example of spatially synchronized life stage phenology in nature. Within regions (``broods'') spanning 50,000 to 500,000 km$^2$, adults emerge synchronously every 13 or 17 years. While satiation of avian predators is believed to be a key component of the ability of these populations to reach high densities, it is not clear why populations at a single location remain entirely synchronized. We develop nonlinear Leslie matrix-type models of periodical cicadas that include predation-driven Allee effects and competition in addition to reproduction and survival. Using both analytical and numerical techniques, we demonstrate the observed presence of a single brood critically depends on the relationship between fecundity, competition, and predation. We analyze the single-brood, two-brood and all-brood equilibria in the large life-span limit using a tractable hybrid approximation to the Leslie matrix model with continuous time competition in between discrete  reproduction events.  Within the hybrid model we prove that the single-brood equilibrium \textcolor{black}{is the only stable equilibrium}. This hybrid model  allows us to quantitatively predict  population sizes and the range of parameters for which \textcolor{black}{the stable single-brood and unstable two-brood and all-brood equilibria exist}.  The hybrid model  yields a good approximation to the numerical results for the Leslie matrix model for the biologically relevant case of a 17-year lifespan. 

\end{abstract}

\maketitle

\section{Introduction}

Periodical cicadas, {\it Magicicada} spp., are remarkable insects in many ways. These species are characterized by exceptionally long life cycles (either 13 or 17 years), their body size is large and they exist at very high densities, sometimes exceeding 500/m$^2$ (\cite{Williams1995}). Another fascinating aspect of these species is that the developmental timing of populations are entirely synchronized; at any one location, periodical cicadas live the majority of their lives underground as nymphs feeding on roots and then emerge as adults in a single year (\cite{Leonard1964,White1975,Williams1995}). The timing of adult emergence is synchronized across large geographical areas ranging from 50,000 to 500,000 km$^2$. Cohorts of developmentally synchronized individuals are called ``broods'' and while their synchronized emergence is believed to be related to predator satiation, it is not clear why only a single brood is normally present in a given spatial location (\cite{Lloyd1966,Dybas1974,Williams1995}). Here we use a Leslie matrix model together with a novel hybrid approximation to explore the cause of synchronized cicada emergence.

Though periodical cicadas are exceptional in many ways, aspects of their life cycle are shared by a larger category of periodical insects. A species is ``periodical'' if its life cycle has a fixed length of several years with adults appearing synchronously in a single year (\cite{Bulmer1977}). Though the causes of developmental synchrony of periodical insects are often unclear, it has been hypothesized to result either from the impact of predators that numerically respond to variation in prey abundance among years or from asymmetrical competition among individuals of varying age (\cite{Heliovaara1994}).

While several previous studies have used mathematical models to  understand the evolution of the long generation time and prime-numbered life cycle lengths (namely 13 and 17 years) (e.g. \cite{Webb2001,Tanaka2009,Yoshimura2009}), few studies have focused on identifying mechanisms generating developmental synchrony (e.g. \cite{Hoppensteadt1976}). Given that periodical cicadas are semelparous -- only adults reproduce and then die immediately after -- Leslie matrix models are an appropriate framework to mathematically describe their population dynamics. However, Leslie matrix models of periodical cicadas require a $13 \times 13$ or $17 \times 17$ matrix. Additionally, there are several nonlinearities associated with the population dynamics. For example, inter- and intra-brood competition as well as predation-driven Allee effects  (that is, lower densities suffer higher {\it per capita} mortality from predation) are important properties of cicada life cycles (\cite{Karban1982,Williams1993,Gasciogne2004,Koenig2013}). Though there are general results for matrix models of arbitrary dimension (\cite{Davydova2005,Mjolhus2005,Cushing2012}), the analytic tractability of these models is limited by  high dimensionality, especially in the presence of nonlinearities in the population dynamics.

We develop a nonlinear Leslie matrix model of the dynamics of periodical cicadas with a finite number, $q$ of age classes.  To make the analysis tractable, we consider the model in the limit of large $q$ and approximate the Leslie matrix model by a hybrid model in which juveniles evolve in continuous time but adult reproduction is a discrete event (See Chap. 10 of Ref.\ \cite{Deroos2013}).  The hybrid model allows us to replace high-dimensional matrix multiplication in the Leslie matrix model by much simpler differential equation(s). \textcolor{black}{Within the hybrid model we first prove the general result that the only stable equilibrium is the singe-brood equilibrium. We then analyze in detail} three equilibria: the stable single-brood equilibrium, the unstable two-brood equilibrium and the unstable all-brood equilibrium.  
An analysis of the solutions allows us to identify conditions under which these equilibria exist. Finally, we numerically compare our findings to simulations of the model with a finite number of age classes. 

\section{Mathematical Models}

\subsection{\textcolor{black}{Leslie matrix model}}
\label{sec:leslie}
We construct a model of periodical cicadas using a nonlinear Leslie matrix model that incorporates competition and predation-driven Allee effects. Although periodical cicadas are known to have 13 or 17-year lifespans, we generalize so that the lifespan is $q \in \mathbb{N}$ years where $q$ is large. The population density of age-cohort $i$ at year $t$ is given by $x_t^i$ where $i=0,\ldots,q-1$. Hereafter, we refer to all classes $x_t^0,\ldots,x_t^{q-\textcolor{black}{1}}$ as juvenile age classes and $\textcolor{black}{x_t^{\rm e}}$ as the adult age class.
\textcolor{black}{Newly hatched first instar nymphs have density $x_t^0$ and,  hereafter, we refer to individuals in this age class simply as ``first instars.''}
The general structure of a Leslie matrix model of cicada populations with a maximum lifespan of $q$ years is therefore given by
\begin{equation} \label{eq1}
\begin{bmatrix} \displaystyle
 x^0_{t+1}\\
x^1_{t+1} \\
x^2_{t+1}\\
\vdots \\
x^{q-1}_{t+1}
\end{bmatrix} =
\begin{bmatrix}
0 & 0 & \ldots & 0 & {\color{black} \R (x_t^{\rm e}) s_{q-1}(\vec{x}_t) }\\
s_0(\vec{x}_t)  & 0 & \ldots & 0 & 0\\
0  & s_1(\vec{x}_t) & \ldots & 0 & 0 \\
\vdots & \vdots   & \ddots & \vdots  & \vdots\\
0   & 0  &\ldots  &  s_{q-2}(\vec{x}_t) & 0
\end{bmatrix}
\begin{bmatrix}
 x^0_t\\
x^1_t \\
x^2_t\\
\vdots \\
x^{q-1}_t
\end{bmatrix} .
\end{equation}
Here, $s_i(\vec{x}_t)$ captures the survivorship of individuals of age $i$ to age $i+1$ where $\vec{x}_t$ is a vector of the population densities for each age class. We assume that survivorship depends on both density-independent (mortality) and density-dependent (competition) processes. We also assume that competition occurs among all juvenile age classes and for analytic tractability we assume the functional form of competition is linear.  Additionally, $\R (x_t^{\rm e})$ is the overall fecundity of the adult population in the presence of Allee effects. We assume that competition occurs prior to the emergence of adults; therefore, $x^{\rm e}_t= s_{q-1}(\vec{x}_t)x^{q-1}_t$. Now, for each age cohort $0 \leq i \leq q-1$, the yearly survivorship is
\begin{equation} \label{eq2}
s_{i}(\vec{x}_t) = \max\Bigg[0,\left(1- \frac{\s}{q} \right) \left(1- \frac{\beta}{q} \sum_{j=0}^{{\color{black} q-1}} x_t^j\right) \Bigg],
\end{equation}
where $\s$ determines mortality and $\beta$ controls the competition of juveniles. The maximum ensures non-negative population values. For simplicity, we assume that $\s$ and $\beta$ do not vary across age classes. Note that both parameters $\s$ and $\beta$ are divided by $q$ to ensure that the large $q$ limit is well-defined while holding these parameters fixed. Also, for fixed $q$ we require that $\s<q$.

The population dynamics described by 
this Leslie matrix model can alternatively be written as $q+1$ equations, one equation describing reproduction,
\begin{equation}
\label{eq:adult}
x_{t+1}^0= \R ( {\color{black} x^{\rm e}_t ) x^{\rm e}_t} ,
\end{equation}
where $x_{t+1}^0$ is the number of offspring produced by $\textcolor{black}{x_t^{\rm e}}$ at time $t+1$, and $q$ equations describing survival,
\begin{equation}\label{eq:junior}
x_{t+1}^{i+1} = \max\Bigg[0, x_t^i \bigg(1-\frac{\s}{q}  \bigg)  \bigg(1 - \frac{\beta}{q} \sum_{j=0}^{{\color{black}q-1}} x_t^j  \bigg)  \Bigg] ,
\end{equation}
for $0 \leq i < q-1$ where here $x_{t+1}^q=x^{\rm e}_{t+1}$. A given age cohort of individuals is referred to as a brood; more precisely, a brood is a group of individuals that all emerge during the same year and brood $k$ reaches adulthood at times  $t \equiv k \pmod{q}$. The term ``brood''  is used throughout the periodical cicada literature though ``year class" is sometimes used to describe such cohorts in other semelparous organisms.

\textcolor{black}{The reproduction factor} $\R$ accounts for fecundity and predation-driven Allee effects. 
We assume that predation occurs during the time interval in which cicadas mate and produce eggs. Additionally, predator satiation occurs at sufficiently high densities of cicadas. Therefore, the functional response mimics that of a Type II Holling response in discrete time:
\begin{equation}
\label{eq:R}
\R({\color{black} x^{\rm e}_t})=\max \bigg[ 0, m  \bigg(1 - \frac{\Pm}{\Ah+{\color{black} x^{\rm e}_t}} \bigg) \bigg],
\end{equation}
where $\Pm/\Ah >0$ is the maximum density of cicadas predated per year and $\Ah >0$ is the population density at which half of the maximum predation is achieved. \textcolor{black}{The reproduction factor incorporates positive density dependence, i.e.\ $R^\prime({\color{black} x)}\geq 0$.}  We assume that $\Pm \geq \Ah$ so that sufficiently low adult population densities do not survive to reproduce.  Note that $m$ also includes adult survivorship.  

\color{black}
\subsection{The hybrid model}
\label{sec:hybrid}
In this section, we present a hybrid model, formally derived  from the Leslie matrix model, Eqns. (\ref{eq:adult})-(\ref{eq:junior}) by taking the limit of large $q$ assuming the effect of background mortality and competition is small in a single year. For large $q$ holding $\s$ and $\beta$ fixed, we can ignore higher order terms in $1/q$ and the difference equations (Eqn.\ (\ref{eq:junior})) become differential equations for the density $x(j,\tau)$ of brood $j$ as a function of a scaled real time variable $\tau=t/q$.  The $q$ differential equations describing the dynamics of the broods between reproduction are
\begin{equation}
\label{eq:juniorh}
\frac{\partial x(j,\tau)}{\partial \tau} = -\s x(j,\tau) - \beta x(j,\tau) \G(\tau),
\end{equation}
where $\G(\tau)$ is the total juvenile population at time $\tau$:
\begin{equation}
\label{eq:G}
\G(\tau) = \sum_{k=0}^{q-1} x(k,\tau).
\end{equation}
In the hybrid model, reproduction is taken to be an instantaneous event occurring simultaneously for every individual in a brood so the population density variable, $x(j,\tau)$ is discontinuous at the reproduction times $\tau_j$ of brood $j$.  We define $\tau_j$ to be any of the reproduction times of brood $j$, which satisfy
\begin{equation}
\label{eq:Rtime}
\tau_j - \lfloor\tau_j\rfloor= j/q.
\end{equation}
For any $\tau_j$, $x(k,\tau_j^-)$ 
is the population density of adults 
while 
$x(k,\tau_j^+)$ is the first instar population density of brood $j$ at times $\tau_j$.  
Equation (\ref{eq:adult}) becomes, for  brood $j$ at any of its reproduction times $\tau_j$,
\begin{equation}
\label{eq:adulth}
x(j,\tau_j^+)= \R \big[ x(j,\tau_j^-)\big]x(j,\tau_j^-).
\end{equation}
Finally, these equations must be supplemented with absorbing state conditions represented by the zero in Eqn.\ (\ref{eq:junior}) so that if $x(j,\tau)$  becomes zero or negative at any time it is thereafter always zero.

\section{General properties of steady states}

Let $\uu_{ij}(\tau)$ be the ratio of the population densities of broods $i$ and $j$ at time $\tau$, $\uu_{ij}(\tau)=x(i,\tau)/x(j,\tau)$.  It is straightforward to show from Eqn.\ (\ref{eq:juniorh}) that $$\dfrac{d \uu_{ij}(\tau)}{d\tau}=0 ,$$ 
so that $\uu_{ij}(\tau)$ is stepwise constant and changes only at times when broods $i$ or $j$ reproduce.  This stepwise constant feature holds also for the original Leslie model and follows from the simplification that all broods experience the same net competition and the same mortality independent of their age class.  The discontinuity at a reproduction time, $\tau_i$ in $\uu_{ij}(\tau_i)$ is given by 
\begin{equation}
\label{}
\uu_{ij}(\tau_i^+) = \R\big[ x(i,\tau_i^-)\big] \uu_{ij}(\tau_i^-),
\end{equation}
and a similar expression holds for the jump at $\tau_j$ in $\uu_{ij}(\tau_j)$ except that the reproduction factor appears in the denominator.  Thus the change in $\uu_{ij}$ from one generation to the next is given by
\begin{equation}
\uu_{ij}(\tau+1) = \frac{\R\big[ x(i,\tau_i^-)\big]}{\R\big[ x(j,\tau_j^-)\big]} \uu_{ij}(\tau),
\end{equation}
with both reproduction times $\tau_i$ and $\tau_j$ chosen to be in the range $\tau$ to $\tau+1$.  For any steady state and for all pairs $i$ and $j$ we require that $\uu_{ij}(\tau+1)=\uu_{ij}(\tau)$ so, for extant pairs of broods, the ratio of their reproduction factors must be equal.  Since $\R$ is single-valued, we have the following result.
\begin{theorem}
\label{thm:1}
In all steady states, all broods with nonzero population density have the same population density as first instars or, equivalently, as adults. 
\end{theorem}

We now show that steady states with more than one extant brood are linearly unstable. Let $\bar{x}(j,\tau)$ be the steady state population density of brood $j$ at time $\tau$ and let $\bar{x}_{\pm} = \bar{x}(j,\tau_j^\pm)$ be the (brood-independent) steady state adult ($-$) and first instar ($+$) population densities, respectively.   Suppose that at some time, $\tau_{\rm in}$, broods are perturbed by a small amounts $\epsilon(j,\tau_{\rm in})$ relative to their  steady state values at $\tau_{\rm in}$ and that generally for $\tau \geq \tau_{\rm in}$ we define $\epsilon(j,\tau)$ from the following:
\begin{equation}
x(j,\tau) = \bar{x}(j,\tau)(1 + \epsilon(j,\tau)).
\end{equation}
Furthermore, suppose that at $\tau_{\rm in}$ at least one brood has a population density greater than its steady state value and at least one brood has a population density less than its steady state value.
Let $\lam(\tau)$ be the difference in the relative perturbation of the most positively and most negatively perturbed brood at time $\tau$:
\begin{equation}
\lam(\tau)=\emax(\tau)-\emin(\tau),
\end{equation}
where $\emax(\tau)=\max_j[\epsilon(j,\tau)]$ and $\emin(\tau)=\min_j[\epsilon(j,\tau)]$.  The maximum perturbation magnitude, $\epsilon^*(\tau)$ is defined as $$\epsilon^*(\tau) = \max[\emax(\tau),|\emin(\tau)|].$$
In the case that $\tau$ is a reproduction time, $\emax(\tau)$ and $\emin(\tau)$ could be discontinuous and it may be necessary to specify $\tau^+$ or $\tau^-$.  

\begin{theorem}
\label{thm:2}
If $\emax(\tau_{\rm in}) > 0$ and $\emin(\tau_{\rm in}) < 0$ then, to linear order in $\epsilon^*(\tau)$ and for  $\tau>\tau_{\rm in}$,  $\lam(\tau)$  is a stepwise increasing function of $\tau$ and  $\lam(\tau +1) > \lam(\tau)$.  
\end{theorem}

The linearized equation of motion for $\epsilon(j,\tau)$ between reproduction events follows immediately from Eqn.\ (\ref{eq:juniorh}) and takes the brood-independent form:
\begin{equation}
\frac{\partial \epsilon(j,\tau)}{\partial \tau} = - \beta \delta \G(\tau),
\end{equation}
where,
\begin{equation}
\label{eq:dG}
 \delta \G(\tau) = \sum_{i=0}^{q-1}\bar{x}(i,\tau)\epsilon(i,\tau) .
\end{equation}
Since all perturbations evolve in the same way, between reproduction events $\lam(\tau)$ is constant. 

Next we show that $\emax(\tau)> 0$ and $\emin(\tau) < 0$ for all $\tau>\tau_{\rm in}$.  First consider times for which there are no reproduction events.  Since there are $q$ broods, $\bar{x}_+ >\bar{x}(i,\tau)$ and $\epsilon_{\rm max}(\tau) \geq  \epsilon(i,\tau)$,
we have from the definition (\ref{eq:dG})  that $q \bar{x}_{+} \emax(\tau)  \geq \delta \G(\tau)$  so that 
$$\frac{d \emax(\tau)}{d \tau} \geq - \beta q \bar{x}_{+} \emax(\tau)$$
If this inequality is replaced by an equality, then the solution is an exponential and is always positive. Thus $\emax(\tau)$ is always positive.  Similarly $q \bar{x}_{+}\emin(\tau) \leq -\delta \G(\tau)$ and  $\emin(\tau)$ is always negative. 

When a brood reproduces the relative magnitude of its perturbation from the steady state increases but does not change sign.  To see this expand the reproduction equation (\ref{eq:adulth}) to linear order in $|\epsilon(j,\tau_j^-)|\leq \epsilon^*(\tau_j^-)$, with the result that
\begin{equation}
\epsilon(j,\tau_j^+) = \Lam \epsilon(j,\tau_j^-),
\end{equation}
where
\begin{equation}
\Lam = 1 + \bar{x}_-  \R^\prime(\bar{x}_-)/R(\bar{x}_-).
\end{equation}
From  positive density dependence  $\Lam >1$ so that $\epsilon(j,\tau_j^+)/\epsilon(j,\tau_j^-) > 1$. Since $\emax(\tau)$ and $\emin(\tau)$ are, respectively, the maximum and minimum values of $\epsilon(j,\tau)$, it is also the case that $\emax$ and $\emin$ do not change sign at the reproduction times of any broods.  Thus $\emax$ and $\emin$ retain their original signs for all times.  Furthermore, in the generation from $\tau$ to $\tau+ 1$ both $\emax$ and $|\emin|$ must increase.  If the brood associated with $\emax$ does not change during the generation then its reproduction event causes an increase in $\emax$.  If a new brood takes over the lead during the generation, it occurs at its time of reproduction and $\emax$ increases at that time.  A similar argument holds for $|\emin|$.  Thus $\lam$ is stepwise increasing and must increase during a generation (one unit of scaled time). The conclusion is that all multi-brood steady states are linearly unstable.

It is noteworthy that the results in this section depend only on positive density dependence and on competition and juvenile mortality that are the same for all age classes.  Thus Theorems 1 and 2 apply to a broad range of competition and reproduction laws.  The linear instability of multi-brood states has been recently proved under certain conditions for the original Leslie matrix model (\cite{Diekmann2018}).

\color{black}

In the sections that follow we develop this hybrid description and  explicitly solve the model for several steady states. Based on these findings, we perform numerical tests to determine how well the hybrid model approximates Leslie matrix model for the realistic case, $q=17$.

\section{Single-brood equilibria for large $q$}
\label{sec:minfty}

In this section, we analytically determine the single-brood stationary state for the model described in Eqns.\ \textcolor{black}{(\ref{eq:juniorh})-(\ref{eq:adulth})}. 
The differential equation describing the age-dependent dynamics of the population density of the single extant brood, $x(\tau)$ is
\begin{equation}
\label{eq:junior1d}
\frac{d x(\tau)}{d \tau} = -\s x(\tau) - \beta x(\tau)^2.
\end{equation}
\textcolor{black}{Let the extant brood be brood zero, and let $\bar{x}_+=\bar{x}(0,\tau_0^+)$  be the equilibrium first instar density,   $\bar{x}_-=\bar{x}(0,\tau_0^-)$ the equilibrium adult density, and  $\bar{x}(\tau)=\bar{x}(0,\tau)$ the equilibrium population density at scaled times $\tau$ between reproduction events.} 
The solution to the differential equation (\ref{eq:junior1d}) is
\begin{equation}
\label{eq:cohortsize}
\frac{1}{\textcolor{black}{\bar{x}(\tau)}} =  \frac{e^{\s \tau}}{\textcolor{black}{\bar{x}_+}}+ \frac{\beta}{\s} (e^{\s \tau} -1),
\end{equation}
and the adult density can be found explicitly as
\begin{equation}
\label{eq:cohortsize1}
\textcolor{black}{\bar{x}_-} =  \left( \frac{e^{\s}}{\textcolor{black}{\bar{x}_+}}+ \frac{\beta}{\s} (e^{\s } -1) \right)^{-1}.
\end{equation}

\textcolor{black}{The adult and first instar population densities are also be related by} the reproduction equation (\ref{eq:adult}),
\begin{equation}
\label{eq:newadult}
\textcolor{black}{\bar{x}_+} = \R \big[\textcolor{black}{\bar{x}_-}\big] \textcolor{black}{\bar{x}_-}.
\end{equation}
Inverting Eqn.\ (\ref{eq:cohortsize1}) to obtain the first instar density as a function of the future adult density and combining the result with (\ref{eq:newadult}) yields a difference equation for the adult density from one generation to the next. In a steady state these must be equal, yielding an equation of the equilibrium adult population density,
\begin{equation}
\label{eq:generation}
\bigg(\frac{e^{-\s}}{\textcolor{black}{\bar{x}_-}}-\frac{\beta}{\s} (1-e^{-\s} )  \bigg)^{-1} =  \R \big[\textcolor{black}{\bar{x}_-}\big] \textcolor{black}{\bar{x}_-}.
\end{equation}
Since  $x$, $\Pm$, and $\Ah$ in $\R$ (Eqn.\ \ref{eq:R}) are in units of number of cicadas per area while $\beta$ is in units of the inverse of cicadas per area, it is convenient to non-dimensionalize the equations by defining the dimensionless equilibrium adult population,
\begin{equation}
\label{eq:defchi}
\chi = \beta \textcolor{black}{\bar{x}_-} ,
\end{equation}
and dimensionless parameters,
\begin{equation}
\label{ }
\al = \Ah \beta, \mbox{  and   } \de = (\Pm-\Ah) \beta.
\end{equation}

In terms of the dimensionless adult population, $\chi$, we can obtain the population density of each age class using Eqn.\ (\ref{eq:cohortsize}). In particular, the first instar density is given by
 \begin{equation}
\label{eq:sscohortsize}
\textcolor{black}{\bar{x}_+} =  \frac{e^{\s}}{\beta}\bigg(\frac{1}{\chi} - r \bigg)^{-1},
\end{equation}
where $\rs$ is defined as
\begin{equation}
\label{ }
\rs = (e^{\s} - 1)/\s.
\end{equation}
Note that when $\s$ is small, $\rs$ is close to and slightly larger than unity.  The effect of mortality can be absorbed in the other parameters by defining an effective fecundity $\m$,
\begin{equation}
\label{eq:effm}
\m = m e^{-\s} ,
\end{equation}
and scaling the other parameters and the adult population density by $\rs$,
\begin{align}
\chis&=\rs \chi,\label{eq:chis}\\
\als&=\rs \alpha,\label{eq:als}\\
\des&=\rs \delta \label{eq:des} .
\end{align}
Note that the positivity of $\Ah$ implies that $\als >0$ and our assumption that $\Pm \geq \Ah$ implies that $\des \geq 0$.  In terms of the dimensionless adult population and dimensionless parameters, the \textcolor{black}{steady state condition, Eqn.\ (\ref{eq:generation}) together with the reproduction equation \eqref{eq:R}, yields the following equation}:
\begin{equation}
\label{eq:eq}
\bigg(\frac{1}{\chis}-1   \bigg)^{-1} = \m \chis \bigg(\frac{\chis - \des}{\chis + \als}\bigg),
\end{equation}
\textcolor{black}{or, written in the standard form of a quadratic equation,}
\begin{equation}
\label{eq:quadratic}
\m^2 \chis^2  +  \m\big(1-\m (1+  \des ) \big)\chis  + \m (\als + \m \des)=0.
\end{equation}
The explicit equilibrium solutions are $\chis=0$ and the two non-trivial solutions $\chis^\pm$,
\begin{equation}
\label{eq:fullsoln}
\chis^\pm = \frac{1}{2} \bigg[ 1 + \des  -1/\m  \pm \sqrt{(1 + \des -1/\m)^2 - 4 (\als/\m + \des)} \bigg].
\end{equation}

\section{Existence, Stability and Properties of the Single-Brood Solutions}
Single-brood solutions $\chis^\pm$ are real if the discriminant, ${\cal D}$, in Eqn.\ (\ref{eq:fullsoln}) is non-negative,
\begin{equation}
\label{eq:D}
{\cal D} = (1 + \des -1/\m)^2 - 4 (\als/\m + \des) \geq 0.
\end{equation}
Since ${\cal D}$ is quadratic in $\des$, opens upward and is negative if $\des = (m+1)/m$, there are two separated regions where the discriminant is non-negative.  We first consider the biologically relevant case where $\als \leq \m -1$ in which case the two regions where ${\cal D}\geq 0$ are given by
\begin{align}
\label{eq:lower}
\des \leq \frac{\m+1}{\m} - 2\sqrt{\frac{\als + 1}{\m}}, \\
\label{eq:upper}
\des \geq \frac{\m+1}{\m} + 2\sqrt{\frac{\als + 1}{\m}}.
\end{align}
\textcolor{black}{Equation \eqref{eq:eq} can be intrepreted as a dynamical map from the adult population at one generation, $\chis$ on the RHS of the equation, to the next generation, $\chis$ on the LHS.}  Linear stability of the steady state solutions is determined by linearizing this dynamical map.  Letting $\chis = \chis^\pm + \epsilon$, then to linear order in $\epsilon$ we find that
\begin{equation}
\label{eq:linear1}
\epsilon^\prime = (1-\chis^\pm)^2\m\bigg[ 1 - \frac{\als(\als+\des)}{(\chis^\pm + \als)^2} \bigg] \epsilon  \equiv \Lambda^\pm \epsilon,
\end{equation}
\textcolor{black}{where $\epsilon^\prime$ is the size of the perturbation after one generation.}
The factor $(1-\chis^\pm)^2$ is the linearization of the mapping from first instar to adult population while the remaining factors are the linearization of the reproduction equation.
Plugging in the equilibrium solutions (Eqn.\ \ref{eq:fullsoln}), the eigenvalues $\Lambda^\pm$ of the \textcolor{black}{two linearizations} are, respectively,
\begin{equation}
\label{eq:lambda}
\Lambda^\pm =1 - \frac{\pm\sqrt{{\cal D}}  \left[2 \als +\des  (-\des  \m+\m+1)\right]-\des  \m {\cal D}}{2  (\als +\des )}.
\end{equation}
After some complicated algebra carried out by Mathematica (see supplementary materials) one sees that in the lower region, Eqn.\ (\ref{eq:lower}), that $\Lambda^+ < 1$ so that $\chis^+$ is the stable solution while $\chis^-$ is unstable. The proof of the stability of the ``$+$" solution and instability of the ``$-$" solution is much simpler in the case $\des=0$, a value that is always in the lower region.  In this case $\Lambda^\pm = 1 \mp \sqrt{{\cal D}}$ so $\Lambda^+$ is less than unity and vice versa.  By  contrast, in the upper region, Eqn.\ (\ref{eq:upper}), the $-$ solution is stable.  However,  the analysis thus far does not incorporate the possibility that the zero is invoked in reproduction equation (Eqn.\ \ref{eq:R} ).   At the edge of stability of the solutions (${\cal D}=0$)  we see from Eqn.\ (\ref{eq:fullsoln}) that
 \begin{equation}
\label{eq:disc}
\chis^\pm=\chis_c=\frac{1+ \des - 1/\m}{2 }.
\end{equation}
Furthermore, for a given value of $\des$ and $\m$, $\chis^- \leq \chis_c$.  Given a dimensionless adult population, $\chis$, the first instar population is proportional to $\m \chis (\chis - \des)/(\als+\chis)$.  Combined with the upper region bound (Eqn.\ \ref{eq:upper}) we see that the first instar population is always negative so that the zero in Eqn.\ (\ref{eq:R}) must be invoked and the population is extinct.

A similar analysis can be given for the $\als \geq \m -1$ though this regime is not likely to be biologically relevant. In this regime, the upper region is again given by Eqn.\ (\ref{eq:upper}) but the lower region is given by $\des \leq -\als/\m$.  As before, the upper region  corresponds to extinction and, in the lower region, it is the $+$ solution that is stable.

Thus, the stable steady state solution is always $\chis^+$ and, for a fixed value of $\des$ and $\m$,  $\chis^+ \geq \chis_c$.  Furthermore,  if $\des$ is non-negative, $\chis^+ < 2 \chis_c$.    As  $\chis$ approaches $\chis_c$ and the limit of linear stability is approached, the  single-brood steady state is increasingly susceptible to extinction from finite perturbations since the ``$+$'' and ``$-$'' solutions of the quadratic equation become closer and if the population falls below the ``$-$'' solution, the flow is toward extinction.

To summarize, a stable single-brood steady state given by $\chis^+$ exists whenever the discriminant  is non-negative, ${\cal D} \geq 0$, and the bounds $\m >1$, $0 < \als \leq \m-1$ and $-\als \leq \des \leq (\m+1)/\m  - 2\sqrt{(\als + 1)/\m}$ are all satisfied.  Since $\des \leq 1$, the steady-state dimensionless adult population size, $\chis^+$ is \textcolor{black}{less than one}.  Of course, the actual equilibrium adult population density $\textcolor{black}{\bar{x}_-}$ may be large if competition is weak since $\textcolor{black}{\bar{x}_-}=\chis^+/\beta \rs$.


\section{Simplified functional response}
\label{sec:simpler}

In addition to considering a Type II Holling response for predation, Eqn.\ (\ref{eq:R}), we consider a limiting case that simplifies the solutions for $\chis$. In particular, we consider a functional response such that the reproductive rate of the population increases linearly as the population density of cicadas increases, thus imposing a weak predation-driven Allee effect. This is achieved by letting $\des =0$, with both $\m \rightarrow \infty$ and $\als \rightarrow \infty$ while holding $\kas =  \als/ \m$ fixed.  In this regime, the total number of offspring is of the form $\R[x] x = (m/\Ah) x^2= (\beta e^{\s}/\kas) x^2$. Here, the stable solution simplifies to the form
\begin{equation}
\label{eq:quad}
\chis = \frac{1}{2} \bigg( 1 + \sqrt{1 - 4  \kas  } \bigg).
\end{equation}
Note that the single-brood solution exists only for
\begin{equation}
\label{relationship}
 \kas \leq 1/4.
\end{equation}

In contrast,  only the extinct solution, $\chis=0$, exists for  $ \kas> 1/4$. If $ \kas = 1/4$ then $\chis=1/2$ or $\textcolor{black}{\bar{x}_-}=1/(2 \beta \rs)$ whereas the first instar population is given by $\textcolor{black}{\bar{x}_+} =e^{\s}/\beta \rs$ (from Eqn.\ \ref{eq:sscohortsize}).  Therefore, there is a discontinuous transition to the single-brood state at $ \kas =1/4$. Finally, for small $\kas$ we find that $\textcolor{black}{\bar{x}_-} \rightarrow 1/ \beta \rs$ and $\textcolor{black}{\bar{x}_+} \sim e^{\s} / \kas \beta \rs $.  As we shall see, the simplified functional response analyzed in this section yields results for the single-brood equilibria that are qualitatively similar to the more general situation when $\m$, $\als$ and $\des$ are all finite.  In Sec.\ \ref{sec:allminfty} and \ref{sec:quadnum}  we also make use of the related parameter $\ka = \al/\m$.

\section{Two-brood Equilibria}
\label{sec:2brood}
In this section we investigate the existence and stability of two-brood equilibria in the setting of the hybrid model.  Let $x(\tau)$ and  $y(\tau)$ be the populations of the two broods at scaled time $\tau = t/q$.  If we assume that inter- and intra-brood competition are equal, then the generalization of Eqn.\ (\ref{eq:junior1d}) to two broods is the pair of equations, 
\begin{align}
\label{eq:2x}
\frac{d x(\tau)}{d \tau} &= -\s x(\tau) - \beta x(\tau)\left[x(\tau) + y(\tau)\right],  \\
\label{eq:2y}
\frac{d y(\tau)}{d \tau} &= -\s y(\tau) - \beta y(\tau)\left[y(\tau) + x(\tau)\right].
\end{align}
Though we assume that the two broods compete with one another as juveniles underground, they emerge at separate times so that the reproduction equation, Eqn.\ (\ref{eq:newadult}) with ${\cal R}$ defined in Eqn.\ (\ref{eq:R}), applies independently to each brood.  Eqns.\ (\ref{eq:2x}) and (\ref{eq:2y}) \textcolor{black}{together with the reproduction events at two times} present mathematical challenges. However, the situation that the two broods emerge at nearly the same time can be addressed using similar methods to the single-brood case.  This symmetric limit where the time between emergences goes to zero should be approximately correct for the case that the two broods are separated by only a few years.  We \textcolor{black}{conjecture} this limit also gives qualitatively correct results even when the time difference between brood emergences is not small.   It should be stressed that the symmetric limit still includes the assumption that the two broods emerge and reproduce independently.

The symmetric limit is invariant with respect to the exchange of $x$ and $y$ so the natural variables for analyzing the problem are sum and difference variables.  Let $z(\tau) = x(\tau) + y(\tau)$ and $w(\tau) = x(\tau) - y(\tau)$.  The equations for $z$ and $w$ are
\begin{align}
\label{eq:z}
\frac{d z(\tau)}{d \tau} &= -\s z(\tau) - \beta z(\tau)^2 , \\
\label{eq:w}
\frac{d w(\tau)}{d \tau} &= -\s w(\tau) - \beta z(\tau)w(\tau) .
\end{align}
The equation for the sum variable $z$ is identical to Eqn.\ (\ref{eq:junior1d}) for a single brood.   
\textcolor{black}{Theorem \ref{thm:1} shows that} the two-brood equilibrium must have $x=y=z/2$ and $w=0$.  Let $\chiz$ be the dimensionless {\it total} equilibrium adult population, $\chiz = \beta r \textcolor{black}{\bar{z}_+}$.   The equilibrium equation for either brood takes a form similar to the single-brood equilibrium equation (Eqn.\ \ref{eq:eq}),
\begin{equation}
\label{eq:eq2}
\frac{1}{2}\bigg(\frac{1}{\chiz}-1   \bigg)^{-1} = \m \frac{\chiz}{2} \bigg(\frac{\chiz/2 - \des}{\chiz/2 + \als}\bigg).
\end{equation}
 After manipulating factors of two, we see that this equation is identical to the single-brood equilibrium equation (Eqn.\ \ref{eq:eq}) except that $\als \rightarrow 2 \als$ and $\des \rightarrow 2 \des$ so that
 \begin{equation}
 \label{eq:soln2}
 \chiz(\m, \als, \des)=\chis^+(\m, 2\als, 2\des),
 \end{equation}
where $\chis^+$ is the dimensionless adult population in the singe-brood equilibrium (Eqn.\ \ref{eq:fullsoln}).
An interesting conclusion is that if all the parameters but the competition are held fixed, then the adult population density of either of the two broods is exactly 1/4  of what it would be if there was only a single brood and the competition was doubled.    The limit of existence of the two-brood equilibrium also extends to only half the competition level as the one-brood equilibrium.  Theorem \ref{thm:2} shows that the two-brood equilibrium is unstable.

\section{All-brood equilibria}
\label{sec:allminfty}

We have seen in the previous sections that for some range of parameters, it is possible to have stable single-brood equilibria and unstable two-brood equilibria. In this section we consider  all-brood equilibria. In these equilibria, all age cohorts are present \textcolor{black}{ and it is a corollary to Theorem \ref{thm:1} that } each brood's size is the same function of age.  We again assume that inter- and intra-brood competition are equal.  We give a general equation for these equilibria. 

\color{black}
In the symmetric all-brood steady state\textcolor{black}{, for $q \rightarrow \infty$, the total juvenile population density $\G$ (Eqn.\ \ref{eq:G}) does not fluctuate in time, $\G(\tau) \rightarrow \G$} 
a constant. 
The differential equation for the population density $x(\tau)$ of any of the equivalent broods is 
\begin{equation}
\label{eq:junior1a}
\frac{d x(\tau)}{d \tau} = -\big(\s + \beta \G \big) x(\tau).
\end{equation}
\textcolor{black}{For the case of a brood that reproduces at time zero,} the solution to this equation is
\begin{equation}
\label{eq:allsoln}
\textcolor{black}{\bar{x}}(\tau)= \textcolor{black}{\bar{x}_+} e^{ -(\s + \beta  \G) \tau}.
\end{equation}
In the large $q$ limit, the sum defining $\G$ can be replaced by an integral \textcolor{black}{of the population of a single brood over one generation between reproduction events},
\begin{equation}
\label{eq:allint}
\G= q \int_0^1  \textcolor{black}{\bar{x}}(\tau) d\tau ,
\end{equation}
and, combining Eqns.\ (\ref{eq:allsoln}) and (\ref{eq:allint}), we find an integral equation for the stationary solution for $\G$,
\begin{equation}
\label{ }
\G = q \textcolor{black}{\bar{x}_+} \int_0^1  e^{ -(\s + \beta  \G) \tau} d\tau .
\end{equation}
Carrying out the integral and using (\ref{eq:allsoln}) to replace $\textcolor{black}{\bar{x}_+}$ by $\textcolor{black}{\bar{x}_-}$ yields the following equation,
\begin{equation}
\label{eq:G}
q \textcolor{black}{\bar{x}_-} \big( e^{\s + \beta  \G}-1 \big) = (\s + \beta \G) \G.
\end{equation}
or, in terms of dimensionless variables,
\begin{equation}
\label{eq:chig}
q \chi \big( e^{ \s + \g}-1 \big)= (\s + \g) \g ,
\end{equation}
where $\g= \beta \G$ and, as before, $\chi = \beta \textcolor{black}{\bar{x}_-}$.

\textcolor{black}{Setting $\tau=1$ in Eqn.\ \eqref{eq:allsoln} and combining it with the reproduction equation yields the steady state condition for the dimensionless adult population of a single brood,} 
\begin{equation}
\label{eq:allss}
\chi e^{ (\s + \g) }  = \chi \R \big[ \chi /\beta] .
\end{equation}
Using Eqns.\ (\ref{eq:R}), (\ref{eq:chig}),  (\ref{eq:allss}), and the previously defined dimensionless variables, we find a transcendental equation for $\g$,
\begin{equation}
\label{eq:allsol}
e^{ \s+ \g } = m \bigg[ \frac{ \g(\s+\g) - q ( e^{ \s + \g}-1) \de }{\g (\s + \g)+q ( e^{ \s + \g}-1)\al } \bigg] .
\end{equation}

We do not undertake the full analysis of Eqn.\ (\ref{eq:allsol}) but, instead consider the simplified functional response introduced in Sec.\ \ref{sec:simpler}, $\de =0$ and $\m \rightarrow \infty$, $\al \rightarrow \infty$ with  $\ka = \al / \m$ fixed, for which (\ref{eq:allsol}) reduces to the following equation:
\begin{equation}
\label{ }
\ka q e^{\g}(e^{\s+\g} - 1) = \g (\s+ \g).
\end{equation}
We note that combining this equation with Eqn.\ (\ref{eq:chig}) yields the  relation that
\begin{equation}
\label{eq:chik}
\chi = \ka e^{\g},
\end{equation}
valid for the simplified functional response.

If we make the further simplification that $\s=0$, we obtain the following equation:
\begin{equation}
\label{eq:simplesigma}
\ka q e^{  \g} (e^{  \g} - 1) = \g^2.
\end{equation}
We find numerically that Eqn.\ (\ref{eq:simplesigma}) has a non-vanishing solution for $\ka \lesssim 0.244/q$, and, at $\ka \approx 0.244/q$, $\g \approx.63$.  For  $\ka <0.244/q$, $\g$ increases as $\ka$ decreases.  Furthermore, we find that as $\s$ increases, the range of $\ka$ decreases for which a non-vanishing solution exists.  Because of the extra factor of $q$ in the bound for $\ka$ for the all-brood equilibrium compared to the bound $\ka=0.25$ for the single-brood equilibrium, the range of parameters is far narrower and the population of individual broods far smaller for the all-brood equilibria compared to the single-brood equilibrium.  Indeed, in the limit $q \rightarrow \infty$ this equilibrium does not exist.  Comparing the range of existence of the one-, two- and all-brood equilibria we conjecture that generally, $k$-brood equilibria exist for a parameter range that scales as $1/k$.  


Note that in the continuum limit, the same calculation applies to a many-brood equilibrium with $k$ extant broods, $1 \ll k<q$, such that these broods are uniformly distributed between $1$ and $q$.  The only difference is that the population in each extant brood is a factor $q/k$ larger than it would be if all broods were present.

\section{Simulation results for single-brood equilibria}

We performed numerical simulations to assess the agreement between the analytic results of the model in the limit as $q \rightarrow \infty$ and the behavior of the original Leslie matrix model for 17-year periodical cicadas (i.e. $q=17$).   The simulations were carried out by iterating the dynamics of the system in the single-brood state,  Eqns.\ (\ref{eq:adult}) and (\ref{eq:junior}), until the adult population in successive generations is unchanged within a small tolerance.  The initial condition must be chosen in the basin of attraction of the stable single-brood equilibrium solution. If the initial population is less than the unstable solution, the population will go extinct.

In the sections that follow, we first consider the simpler form of the reproduction equation described in Sec.\ \ref{sec:simpler} and Eqn.\ (\ref{eq:quad}) followed by results from the full model with a Type II Holling response, Eqns.\ (\ref{eq:R}) and (\ref{eq:fullsoln}).

\subsection{Simplified functional response}
\label{sec:quadnum}
\begin{figure}[h]
\begin{center}
\includegraphics[scale=0.4]{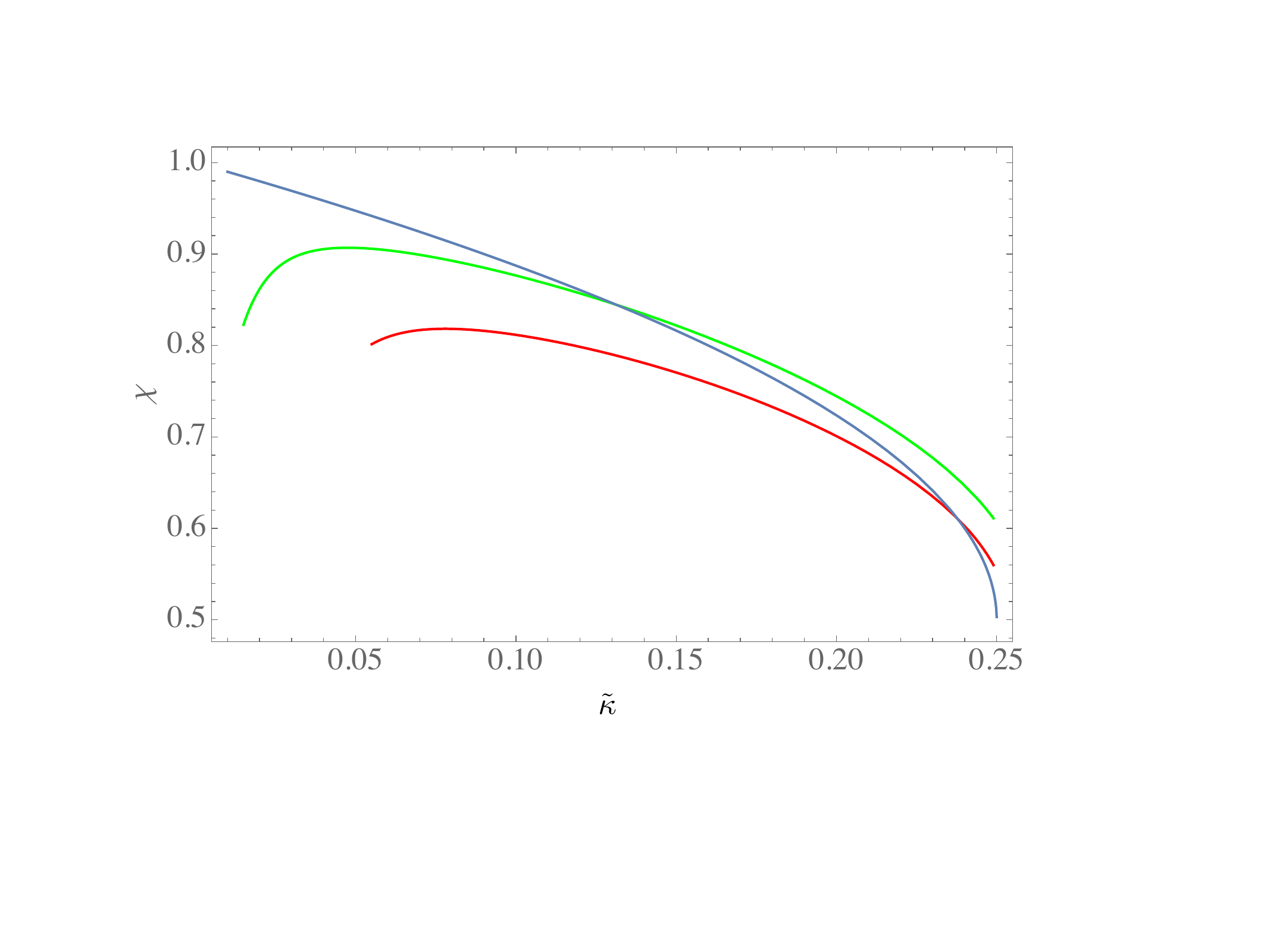}
\caption{
Dimensionless adult population, $\chi$ vs.\ the parameter, $\kas$ for the single-brood stationary state when $m \gg 1$. The blue curve (top for small $\kas$) is the $q\rightarrow \infty$ theoretical result (Eqn.\ \ref{eq:quad}) with $\s=0$, $\de=0$ and $m \rightarrow \infty$ (the simplified functional response). The red  and green curves correspond to the equilibria using  $q=17$, $\s=0$, and $m=1000$. The red curve (bottom for small $\kas$) represents linear competition (Eqn.\ \ref{eq:junior}) while the green curve (middle for small $\kas$) represents exponential competition (Eqn.\ \ref{eq:juniore}).  Period doubling, chaos and extinction for the $q=17$ simulations for small $\kas$ are not shown.  }
\label{fig:th_vs_sim}
\end{center}
\end{figure}

We first consider the simplified functional response discussed in Section \ref{sec:simpler}.   Figure \ref{fig:th_vs_sim} shows the dimensionless adult population $\chi = \beta \textcolor{black}{\bar{x}_-} $ as a function of $\ka= \al / \m$ where $\ka$ is related to the strength of competition and predation relative to fecundity. Note that the simulations are carried out with $\s=0$ so that $\rs=1$, $\ka=\kas$ and $\chi=\chis$.  As seen in Fig. \ref{fig:th_vs_sim}, the behavior of $\chi$ as a function of $\ka$ obtained analytically in the limit $q \rightarrow \infty$, Eqn.\ (\ref{eq:quad})  (blue curve), is in reasonable agreement with  numerical simulations of the Leslie matrix model for $q=17$ and $m=1000$ (red curve) as long as $\ka$ is sufficiently large.  For small $\ka$, however, the $q=17$ simulation result falls below the $q \rightarrow \infty$ solution and then undergoes a period-doubling cascade to chaos (not shown in the figure) within a small range of $\ka$ before going extinct (\cite{Wikan2012}).  At $\ka \approx .044$ the simulations show that there is a bifurcation to a single-brood two-cycle.  In the range $0.0036 \lesssim \ka \lesssim 0.0049$ the behavior is chaotic and for $\ka \lesssim 0.0035$ only the extinct state exists for $q=17$.

For  $q=17$ and $\ka$ small, better agreement with the $q \rightarrow \infty$ theory is obtained using exponential competition,
\begin{equation}
\label{eq:juniore}
x_{t+1}^{i+1} = x_t^i \bigg(1-\frac{\s}{q}  \bigg)  \exp\bigg(-{ \frac{\beta}{q} \sum_{j=0}^{q-\textcolor{black}{1}} x_t^j } \bigg) ,
\end{equation}
which is, in any case, more ecologically realistic but less tractable theoretically.  The green curve in Fig.\ \ref{fig:th_vs_sim} shows the $q=17$,  $m=1000$, $\s=0$ simulation with exponential competition.  The $q=17$ single-brood steady state undergoes a period-doubling transition to chaos in a very narrow range of $\ka$.  Numerical simulations show that the bifurcation to the two-cycle occurs at $\ka \approx .0113$.   In the range $0.038 \lesssim \ka \lesssim 0.040$ the behavior is chaotic and for $\ka \lesssim 0.038$ only the extinct state exists.  Note that for the $q \rightarrow \infty$ the solution exists as $\ka \rightarrow 0$ where $\chi=1$ and period doubling does not occur.  The extinction transition at small $\ka$ for finite $q$ is the result of a population overshoot leading to strong competition in the first instar population and a subsequent crash.

\subsection{Type II Holling response}

\begin{figure}[h]
\begin{center}
\includegraphics[scale=0.5]{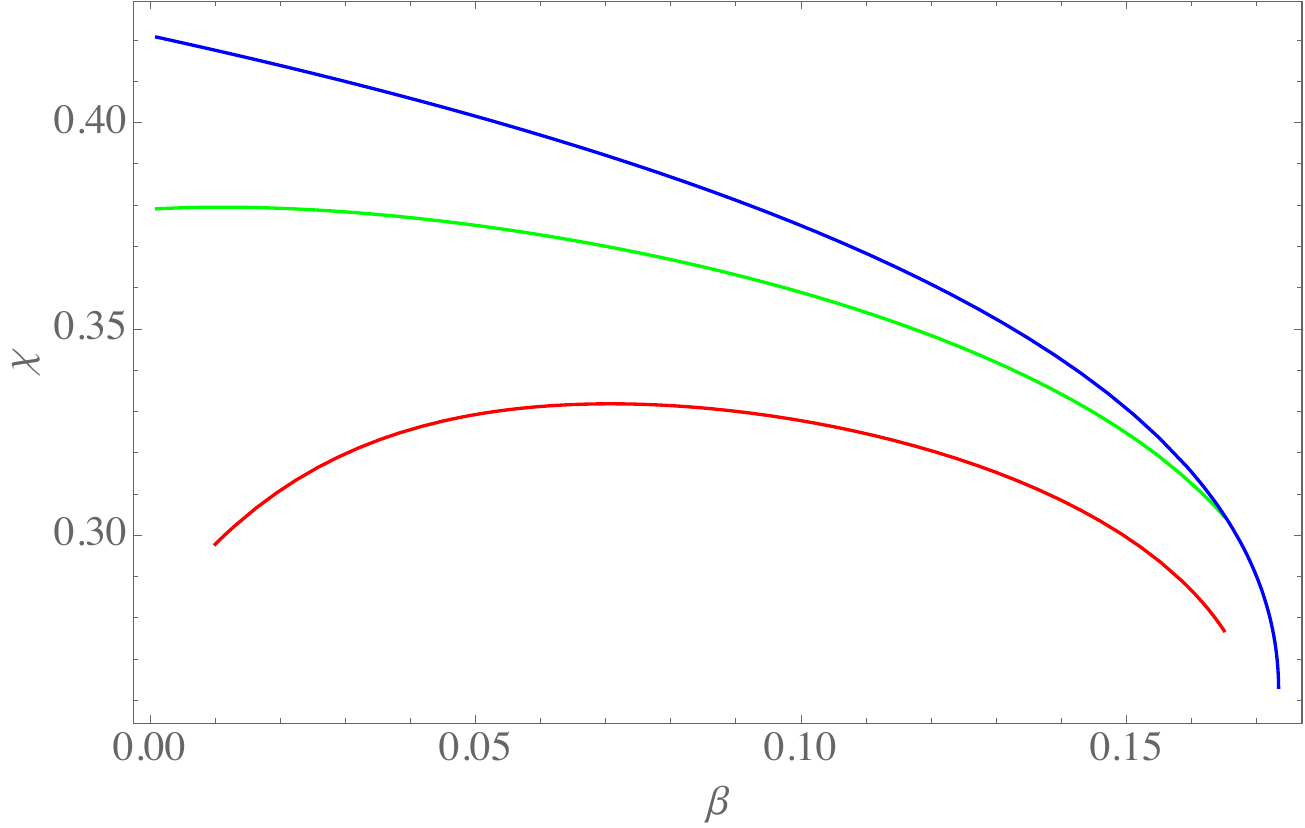}
\caption{Dimensionless adult population, $\chi$ vs.\ the competition parameter, $\beta$ (${\rm m}^2$) for the single-brood stationary state for realistic values of parameters for  17-year cicadas ($m=50$, $\Pm=3.9/$m$^2$, $\Ah=3.3/$m$^2$ and $\s=1.4$). The blue curve (top) is the $q\rightarrow \infty$ theoretical result. The red  and green curves are both  $q=17$, simulations. The red curve (bottom) represents linear competition (Eqn.\ \ref{eq:junior}) while the green curve (middle) represents exponential competition (Eqn.\ \ref{eq:juniore}).  A period-doubling transition for linear competition for $q=17$ occurs for $\beta\approx 0.009$m$^2$ and is not shown.  There is no period-doubling transition for exponential competition.  The edge of stability of the single-brood equilibrium occurs at $\beta_c=0.173$m$^2$. }
\label{fig:th_vs_sim_alex}
\end{center}
\end{figure}

Given that the full version of the model has many parameters, we choose to study a set of parameters that is realistic for 17-year periodical cicadas. Based on the literature reasonable values of the parameters are $m=50$, $\Pm=3.9/$m$^2$, $\Ah=3.3/$m$^2$ and $\s=1.4$ (\cite{inprep}).  Unfortunately, there have been very few studies of competition among cicada nymphs so the nature and magnitude of competition remains an unknown (\cite{Karban1984}). Consequently, here we simply plot varying values of $\chi$ as a function of $\beta$, measured in units of m$^2$.  Figure \ref{fig:th_vs_sim_alex} shows the theoretical  (Eqn. \ref{eq:fullsoln}) $q \rightarrow \infty$ result (blue curve) along with the $q=17$ simulation results for linear competition (red curve) and exponential competition (green curve).  The theory is again a reasonable approximation to the finite $q$ simulations.  In addition, the behavior of the simple (quadratic) functional response, is qualitatively similar to these more realistic parameters.  For the above realistic parameters, we find that the edge of stability of the single-brood steady state obtained from Eqn.\ (\ref{eq:disc}) is at $\beta_c=0.173$m$^2$.
The dimensionless parameters $\al$ and $\de$ vary with $\beta$ but their values at the edge of stability of the single-brood state are as follows:  $\al_c=0.572$ and $\de_c=0.104$.  
The control parameter $\kas=\ka \rs$ at the edge of stability for the simple (quadratic) functional response is $1/4$ whereas for these parameters the edge of stability is $\kas_c = 0.101$.  The quantity $\chis= \chi \rs$ at the edge of stability for the simple quadratic functional response is $1/2$ whereas for these parameters $\chis_c = 0.573$.  Thus realistic parameters for the 17-year cicadas are qualitatively similar to the simplified case of the quadratic functional response.  The critical density of the adult population is, $\textcolor{black}{\bar{x}_-}=\chi_c /\beta_c=1.51/$m$^2$.  Below this density, the single-brood steady state cannot exist.  For weak competition the adult population is inversely proportional to the competition parameter, $\textcolor{black}{\bar{x}_-}\approx (0.4/\beta)$m$^{-2}$. For small $\beta$, the $q=17$ single brood solutions exists in the limit $\beta \rightarrow 0$ for both linear and exponential competition.  For exponential competition there is a unique steady state for small $\beta$.  For linear competition there is a bifurcation at $\beta \approx 0.009$ to a period-2 oscillation followed by a period-doubling cascade to chaos. The population dynamics displays two-banded chaos in the limit  $\beta \rightarrow 0$ for linear competition.  Note that the (unrealistic) $\beta \rightarrow 0$ limit is singular since the population diverges so that linear and exponential competition differ in this limit.

\section{Discussion}

The spatial distribution of periodical cicadas remains enigmatic: all populations exist within  broods spanning large, non-overlapping geographical areas with well-defined boundaries.  Within each brood, development is synchronized such that adults emerge synchronously every 13 or 17 years. 
(\cite{Lloyd1966,Dybas1974,Williams1995}). Using a combination of analytic and numerical methods, we studied a nonlinear Leslie matrix model with the aim of determining the conditions under which a single-brood stable equilibrium exists.  The main mathematical tool employed here is continuous time approximation to juvenile development allowing us to replace the high-dimensional Leslie matrix model by a far more tractable hybrid model. \textcolor{black}{In the context of the hybrid model we proved a theorem showing that all equilibria with more than one extant brood are linearly unstable.  The proof is quite general insofar as it does not depend on the specific forms of competition and reproduction except for the following features.  First, reproduction has positive density dependence and, second, competition applies equally to all juvenile age classes.} The instability of the two-brood and all-brood states arises from the growth of one brood at the expense of the other(s).

Using the hybrid model we studied equilibria consisting of a single brood, two broods and all broods.  We showed that the single-brood equilibrium exists and is stable so long as competition, predation and mortality are not too strong relative to fecundity.  The two-brood and all-brood equilibria exist over a much narrower range of parameters and\textcolor{black}{, according to Theorem \ref{thm:2},} are always unstable.  While we have considered only two multiple-brood equilibria, the methods used here, with additional work, would also be applicable to three and higher numbers of broods equilibria.  The analysis of the two-brood equilibrium in the hybrid model involves two coupled differential equations, while analyzing a $k$-brood state would involve $k$ differential equations. 

Our model contains five parameters that control  competition, the functional response for predation, mortality and fecundity.  We showed that these five parameters can be reduced to three dimensionless parameters that describe the properties of the single-brood and two-brood equilibria, while the all-broods equilibrium requires four dimensionless parameters.  Within the hybrid model, the region of existence and stability of the three equilibria can be easily determined. 

Additional insights into the single-brood equilibrium were obtained by using a simplified quadratic functional response to predation.  Here a single parameter, $\kas$, essentially the ratio of the product of predation and competition to fecundity,  determines the existence of the single-brood steady state.  This ratio must be less than a critical value for the single-brood steady state to exist.  The results for this simplified quadratic functional response are both qualitatively and quantitatively similar to the more general Type II Holling response studied here.  A key insight is that the transition to extinction is discontinuous.  Although the single-brood equilibrium is linearly stable, it becomes increasingly sensitive to large perturbations as the edge of stability is approached and the unstable fixed point approaches the stable fixed point.

The results from the hybrid model were compared to simulations of the Leslie matrix model corresponding to a finite $q$ ($q=17$) -- which is representative of the actual behavior of periodical cicadas -- for both the simple linear form of competition as well as a more realistic exponential functional form. 
We found that the hybrid model is in reasonable agreement with the $q=17$ simulations. However, in some cases the simulations for $q=17$ show a period-doubling cascade followed by extinction if predation and competition are very weak relative to fecundity.  The $q \rightarrow \infty$ limit used in our continuum analysis likely eliminates this more complex behavior. The possibility of period doubling and extinction for the finite $q$ case is intuitively plausible--high fecundity leads to a boom in population size of first instar nymphs, but this amplifies competition and ultimately leads to a crash in the population.  However, this regime is unlikely to be biologically relevant.

While the above observations relied on simplifying the form of predation and assuming a very high fecundity, we also compared our theoretical equilibrium analysis to numerical simulations of the finite $q$ case using parameters realistic for periodical cicadas. Here, the parameters driving fecundity and predation are fixed based on existing data (\cite{Karban1982,Karban1984,Karban1997}), and we only vary the parameter corresponding to the level of competition, reflecting the fact that very little information exists about competition among nymphs. Here, similar observations are made: the single-brood state only stably exists when competition is sufficiently small to avoid driving the population to extinction. Again, we found that the $q \rightarrow \infty$ theory is in reasonable agreement with the $q=17$ simulations.

We made several simplifying assumptions to increase the tractability of the analysis in this paper. For example, we employed a relatively simple functional form for competition.  A recent paper (\cite{inprep})  implements and compares several  alternative functional forms for competition  through numerical simulation of the Leslie matrix model.  This paper also considers the role of ``stragglers''  
--individuals emerging out of sync from their brood as a result of delayed or accelerated development (\cite{White1979a,Lloyd1976,Williams1995,Heath1968,White1979b}). This paper explores the effect of stragglers on the stability a single brood through numerical simulation of the Leslie matrix model. The present results for the unstable two-brood equilibrium are relevant to understanding the dynamics following a large ``leakage'' event.  

Previous studies of periodical insects indicate that the synchronized development characteristic of these organisms may arise either as a result of asymmetrical competition among different age cohorts or as an emergent consequence of numerical responses of predators (\cite{Heliovaara1994}). 
(\cite{Bulmer1977}) found that synchronized development in periodical insects can occur when inter-cohort competition exceeds intra-cohort competition though this seems unlikely for periodical cicadas. Similar mechanisms have also been advanced to explain developmental synchronization in insects that exhibit multiple generations within a single year (\cite{Gurney1983,Hastings1987,Hastings1987b,Hastings1991,Briggs2000,Yamanaka2012}). 
(\cite{Bulmer1977}) concluded that predation by generalist predators may reinforce synchronous development in periodical insects and this is likely the case with periodical cicadas. Hoppensteadt and Keller (\cite{Hoppensteadt1976}) showed that predation and competition can interact to produce synchronous development in cicada populations of lifespans $>10$ years but they only considered intrabrood competition. 
(\cite{Behncke2000}) also explored a model with interacting competition and predation and found that as life span increased, this leads to the emergence of a single synchronized cohort (brood). 

Our work demonstrates how inter-cohort competition together with the Allee effect can drive synchronization in a periodical insect but we did not explore the possible role of numerical responses in predators. For periodical cicadas, there is good evidence for such responses (\cite{Koenig2013}) but their role in synchronization will need to be explored in future studies. 

Periodical cicada broods extend over large regions with sharp boundaries separating broods.  The current study is concerned with a single patch but it would be interesting to extend the methods developed here to a spatially explicit model to study the boundary between broods.  The instability of the two-brood equilibrium helps to explain the sharp boundary between broods. 

\begin{acknowledgements}
The authors thank the Santa Fe Institute for sponsoring three working groups during which much of this work was carried out.  JM, AH and AN acknowledge support from the National Science Foundation under INSPIRE Grant No. 1344187.
We are grateful to Prof. Odo Diekmann for providing key insights that motivated Theorems 1 and 2.
\end{acknowledgements}

\bibliographystyle{spbasic}      

\bibliography{CicadasBib}
\end{document}